\documentclass[12pt,a4paper]{elsarticle}
\usepackage[utf8]{inputenc}
\usepackage[portuguese]{babel}
\usepackage[T1]{fontenc}
\usepackage{amsmath}
\usepackage{amsfonts}
\usepackage{amssymb}
\usepackage[left=2cm,right=2cm,top=2cm,bottom=2cm]{geometry}
\title{Carta a academia: por uma refundação do CNPQ e da ciência brasileira}
\journal{ArXiv}

\begin{document}

\begin{frontmatter}



\title{Carta a academia: por uma refundação do CNPQ e da ciência brasileira}


\author[mymainaddress]{Marcos Paulo Belançon\corref{mycorrespondingauthor}}
\cortext[mycorrespondingauthor]{Corresponding author}
\ead{marcosbelancon@utfpr.edu.br}

\address[mymainaddress]{Universidade Tecnológica Federal do Paraná\\Câmpus Pato Branco -  Grupo de Física de Materiais\\CEP 85503-390, Via do conhecimento Km 01, Pato Branco, Paraná, Brazil}

\begin{abstract}

In this ``Letter'', I do introduce my point of view about the role of science in the development and organization of our societies, emphasizing that science is too far from our societies. In Brazil we follow a way similarly to that of USA and France concerning this phenomenon, marked by the lack of participation of our academies and scientific community in politics and economy, i.e. we can see that strategic decisions are been hold by only electoral bias and, when there is the presence of academic science, usually it is restricted to the fields of economical, political and social sciences. I conclude that the strategic planning of a country, or of the planet as a whole, needs that academy stay close of society, been this one necessary both not enough condition to the proper functioning of any democracy. In the Brazil case more specifically, I propose that this political change in our academies begin by one profound thought about the Brazilian Science with a consequent refoundation of the CNPQ (Nacional Council for Science and Technology development). 

- - - -

Nesta ``carta'', apresento o meu ponto de vista sobre o papel da ciência no desenvolvimento e organização de nossas sociedades, enfatizando que ciência está muito distante da sociedade. No Brasil trilhamos um caminho semelhante ao dos EUA e França no que diz respeito a este fenômeno, marcado pela falta de participação das academias e da comunidade científica na política e economia, i.e. verificamos que decisões estratégicas são tomadas com viés eleitoral e, quando há presença da ciência acadêmica, habitualmente ela se restringe a esfera da ciência econômica, política e social. Concluo que o planejamento estratégico de um país, ou do planeta todo, precisa que a academia se aproxime da sociedade, sendo esta uma condição necessária e não determinante para o bom funcionamento de qualquer democracia. No caso do Brasil mais especificamente, proponho que essa mudança política das academias comesse por uma profunda reflexão sobre a ciência brasileira com a consequente refundação do CNPQ.

\end{abstract}


\end{frontmatter}

A física dirigiu as revoluções científicas nos últimos séculos. Do método científico de Galileu até a difração de elétrons e a equação de Sch\"orodinger; passando pela gravitação de Newton, os raios-x de R\"ontgen  e o méson $\pi$ de César Lattes, entre tantas outras descobertas. A revolução industrial começa com o domínio da máquina a vapor, passa pela exploração do eletromagnetismo; e então a física do estado sólido trás os semicondutores, chegam o laser e as comunicações ópticas. Na sequência de cada um destes desenvolvimentos, vieram muitos avanços na biologia, medicina, e até mesmo na história e arqueologia. A física abriu o caminho para essas e outras áreas da ciência se desenvolverem.

O século XXI começa muito diferente de qualquer século anterior. Os sistemas educacionais do planeta cresceram ainda mais do que a população, de maneira que o acesso as universidades não é mais uma exclusividade de poucos homens e quase nenhuma mulher. A universidade de Zurich tinha pouco mais de 2000 alunos quando Einstein se candidatou a uma vaga, e Marie Curie não tinha uma infinidade de opções já que a universidade de Paris era uma das poucas que ``aceitava'' mulheres no final do século XIX. Hoje a Universidade de São Paulo sozinha possui mais professores do que Zurich tinha como alunos, e no Brasil as mulheres ocupam cerca de metade das vagas universitárias. Entretanto muitos dos problemas deste século deveriam passar por um intenso debate científico; na prática o que continuamos a ver são discussões e tomada de decisões em termos exclusivamente econômicos e políticos.

Num mundo tão diferente do passado, a física talvez seja vítima de seu sucesso. Somos tentados a justificar toda e qualquer pesquisa com a ideia de que é pelo bem comum de todos, como se o financiamento de nossas pesquisas fossem um elemento sagrado do orçamento. Assim dizemos que os governos deveriam investir em pesquisa, mesmo em tempos de crise. Certamente eu acredito que a ciência, sobretudo a brasileira precisa desenvolver-se; entretanto, que tipo de ciência queremos para o nosso país? Nossa sociedade está de fato se servindo de nosso conhecimento? Estamos ensinando o que a população? Vale lembrar que no mundo todo o financiamento da ciência é predominantemente público, e em geral o acesso ao próprio conteúdo produzido em nossas academias não é aberto, i.e. o estado paga para ter acesso aos artigos que financiou.

Vejamos alguns exemplos dos países ditos desenvolvidos que justificam heuristicamente o meu ponto. Todos reconhecemos que os Estados Unidos são o líder mundial em praticamente qualquer parâmetro que almeje medir o desempenho científico; seja pelo número de publicações, de patentes ou de prêmios Nobel. Entretanto, no ano de 2017 ainda é possível que esse país eleja um presidente que ignora toda a sua comunidade científica e faça uma opção na direção dos combustíveis fósseis em detrimento de fontes alternativas ou do investimento no desenvolvimentos de novas fontes para sua matríz. É fato que o sistema educacional americano não se destaca pelo mundo; programas como o ``No Child Left Behind'' foram implementados e pouca ou nenhuma evolução foi observada. Por fim podemos concluir que o mais premiado sistema universitário do planeta não reflete seu desempenho na população do país, que por sua vez não vê um grande problema em declarações absurdas de um presidenciável\footnote{Ele há pouco tempo acusou a administração Obama, que tinha o Nobel Steven Chu como ministro de energia de enfraquecer os EUA ao acreditar na invenção chinesa do aquecimento global.}. Dezenas de milhões de americanos ignoram o que sua academia tem a dizer sobre o assunto.

É fato que o carvão tem sido o combustível responsável pelo crescimento econômico mundial neste século, sobretudo a partir da China. Para cada $MWh$ de energia renovável, $35MWh$ de carvão foram adicionados a matríz mundial nos anos 2000. Enquanto isso, na Europa muitos países começam a se mover na direção das energias renováveis, sem sequer considerar as limitações físicas da questão. Por um lado temos a promessa, principalmente dos ``partidos verdes'' de que o futuro será movido a energia eólica e fotovoltaíca, entretanto as tecnologias que dispomos hoje parecem limitar-se inclusive na disponibilidade de matéria prima em nosso planeta\cite{Grandell2016,Feltrin2007}; e pra mudar o fato de que Prata, Telúrio e Neodímio são raros, precisariamos mudar o ``Big Bang''. E claro, a contribuição das fotovoltaicas tem sido mais a de dar esperança a alguns  países, como a Alemanha, do que de fato resolver o problema; em boa parte da Europa a energia solar produz no inverno $1/5$ de sua capacidade, e isso justamente na época do ano em que o consumo de energia aumenta. O que dizem as academias européias sobre essa questão? Praticamente nada, pois algum barulho vêm das academias apenas para reclamar das medidas de austeridade.

O caso da França e seu programa nuclear é ainda mais interessante. Na década de 1970 se implementou uma opção pela energia nuclear, entre outras coisas convencendo boa parte de sua população de que era a única maneira de construir uma independência energética para o país; na época o mundo era sacudido pela crise do preço do petróleo. Hoje 80\% de sua eletricidade é de origem nuclear, e somando-se a isso o monópolio de tecnologia construído sob a propaganda do tratado de não proliferação, a França consegue participar de projetos de reatores e usinas de reprocessamento por todo mundo. Por este lado o programa parece ter sido um sucesso, entretanto, nem mesmo em território Francês o problema dos resíduos radioativos foi resolvido. A França tem em ``La Hague''\cite{Sanders2016} um dos maiores estoques de lixo a procura de um destino, e muitos defendem que a usina de reprocessamento foi um erro e deveria ser fechada; de lá saem isótopos radioativos gasosos e líquidos que espalham-se pelo norte do país, sem que se saiba quais serão as consequências. 

O programa nuclear francês, sem dar explicações claras a população levou 11 anos e 60 bilhões de francos para construir seu reator ``fast breeder'', chamado de ``Superphènix'', que eliminaria o problema do Plutônio; o reator funcionou por 10 anos e foi desligado permanentemente depois de produzir 2 bilhões de francos de eletricidade. Se não bastasse este episódio, apenas 5 anos depois a França anuncia a criação de uma mega empresa do setor nuclear, a AREVA, que iria unir o ``know-how'' francês para liderar projetos nucleares pacíficos pelo mundo, vendendo produtos e serviços que garantiriam o desenvolvimento econômico. Em 2005 anunciam um contrato para a construção de um reator de terceira geração na Finlândia, que seria o primeiro grande empreendimento da AREVA, no valor de 3 bilhões de Euros. O reator que deveria estar operacional em 2010 continua em construção, e já consumiu 9 bilhões de euros que sairam dos cofres franceses. No começo dos anos 2010, num escândalo de corrupção, a AREVA começou a ser investigada por ter comprado a canadense ``UraMin''; no negócio, dizem os investigadores, o preço da UraMin foi inflado e parte do dinheiro foi utilizado como propina para que magnatas no continente africano garantissem a AREVA contratos para a construção de novos reatores em seus países. 

O domínio da física nuclear tem ainda exemplos mais graves. O projeto Manhattan e a subsequente corrida armamentista deixaram em ``Hanford site'', as margens do rio Columbia,  um dos maiores depósitos de lixo radioativo do planeta em condições precárias. O Reino Unido e outros países despejaram centenas de milhares de toneladas de lixo radioativo no Atlântico. A Rússia aposentou submarinos nucleares com seus reatores no fundo do mar do ártico. O Japão construiu uma usina idêntica a de ``La Hague'', e agora é proprietário de dezenas de toneladas de plutônio a procura de um destino, já que inclusive a maioria dos reatores japoneses que poderiam utilizá-lo está desligada. Outros exemplos semelhantes não faltam.

Qual o papel da ciência nesse tipo de questão? Não somos culpados, mas somos responsáveis por muitos destes problemas; inclusive por nos omitirmos. No caso do Brasil, onde está acontecendo a discussão sobre o destino do lixo de Angra? Onde está acontecendo a discussão sobre novos reatores nucleares? Infelizmente não é nas universidades, tão pouco é no CNPQ; A questão do lixo está nos tribunais, que agora obrigam o estado brasileiro a encontrar um destino definitivo para ele, ao mesmo tempo que os novos reatores devem ser de mais interesse de empreiteiras, investidores de multinacionais do setor nuclear e, por consequência, de partidos políticos. Instituições como o CNEM, a Eletronuclear ou o Ibama não tem como função incluir a sociedade neste debate. As Universidades Públicas do Brasil deveriam cumprir essa função.

Vale lembrar que o nosso CNPQ foi fundado depois da segunda guerra, quando o nosso representante na comissão de energia atômica da recém criada ONU, o Almirante Álvaro Alberto da Motta e Silva, recomendou ao governo a criação de um conselho nacional de pesquisa. E o que tem feito esse conselho nos dias de hoje? Ainda que o CNPQ tenha participado de louváveis feitos, sendo uma das principais agências de fomento e contribuindo portanto para o expressivo aumento do número de doutores no Brasil, é preciso que se discuta o futuro dessa importante instituição.

Ás perguntas que paíram sobre a minha cabeça são, entre outras: O ``conselho nacional de pesquisa'' deve continuar no papel de distribuir uma centena de milhões de reais quantizados em pacotes de 30 ou 50 mil reais? Deve continuar fomentando a numerologia da nossa produção científica? Deve distribuir os ``títulos de nobreza'' na forma de bolsas de produtividade? Isto é o que tinham em mente aquele grupo de militares e cientístas que criaram o projeto científico nacional quando o Brasil tinha 80\% de analfabetismo na década de 1950? 

Convenhamos, essa discussão precisava ter acontecido há décadas; ou melhor, previsava acontecer de maneira contínua. Mas na academia, enquanto o minguado orçamento de ciência e tecnologia não é cortado, sobretudo uma grande parcela da nobreza da produtividade não faz nenhuma questão de discutir a ciência brasileira. Afinal, estão fazendo a parte deles: autores publicam artigos mensalmente, mesmo quando se tem certeza de que às promessas da introdução e da conclusão dos trabalhos é uma fantasia. É necessário que exista pesquisa em ciência básica, sem pretenções imediatas de atingir o público; entretanto, em ``hard science'' é claro que a imensa maioria das pesquisas está desligada da realidade, ou pelo menos da realidade brasileira. 

A busca por indicadores de produção já atingiu os doutorandos, que não tem futuro se não publicarem uma dezena de artigos; atingiu os mestrandos, que num comprimido espaço de 2 anos devem fazer seu mestrado e publicar um artigo. Em muitos casos nem mesmo a publicação de uma dissertação em português é mais necessária, bastando a publicação do artigo em inglês, aumentando ainda mais a distância entre a ciência e a sociedade que a está financiando; por isso tudo não é surpresa encontrar alunos recém graduados desesperados atrás de indicadores de produção que lhe garantam uma vaga de mestrado.

É preciso não diminuir as realizações da ciência brasileira, simbolizadas pela história do CNPQ. Começamos a fazer ciência num contexto de instabilidade política que perdurou da segunda guerra até a década de 1990; aumentamos nossos indicadores, o número de teses defendidas por ano foi multiplicado por 8. Entretanto, o próprio CNPQ reconhece que o investimento privado em pesquisa no Brasil é pífio; o que, convenhamos mais uma vez, não deveria ser uma surpresa: nosso conselho nacional de pesquisa fez o que pôde para não ser engolido em meio a tantos desafios, mas no caminho desfigurou-se em agência de fomento a medida que deixou de dar conselhos ao estado brasileiro e a toda a população sobre que direção devemos tomar.

Não deveria o CNPQ e nossas academias discutirem agora se queremos extrair gás de xisto em nosso território?  Ou se vamos construir uma grande hidrelétrica ou mais alguns reatores nucleares? Não deveríamos estar a frente dessa discussão? Nós não estamos a frente de nenhuma dessas e de tantas outras, porque estamos preocupados com o que o CNPQ e a CAPES querem que seja nossa preocupação. Não deveríamos esperar as usinas do Xingu e de Angra para brigar pela paralisação de um canteiro de obras; deveríamos propor hoje quais serão as usinas que daqui a 20 anos vão garantir a necessidade energética brasileira de maneira que não seja necessário começar um canteiro de uma obra a qual nos opomos. Uma audiência no Senado com meia dúzia de especialistas não é o mesmo que ouvir e debater com a sociedade.

Por todos os motivos justificados nessa carta, manifesto o meu ponto de vista da necessidade de uma refundação da política nacional que vá muito além da ``reforma eleitoral'' em discussão em Brasília. A comunidade academica nacional precisa fazer política, o CNPQ precisa voltar a ser um conselho que oriente políticas estatáis e privadas de desenvolvimento, e nós pesquisadores precisamos voltar a se preocupar com o avanço da ciência, com os problemas e anseios de nossa sociedade. Só precisamos explicar para a sociedade que ela está sendo ameaçada pelos cortes no ministério de ciência e tecnologia porque a sociedade não se sente ameaçada; de fato, o que está ameaçado além do já minguado orçamento do CNPQ são as bolsas de produtividade e os indicadores de produção que a sociedade sequer conhece.

\bibliography{library}

\end{document}